\documentclass[12pt,english, onecolumn, draftcls]{IEEEtran}

\usepackage{graphics, graphicx}
\usepackage{amssymb,amsmath}
\usepackage{amsmath}
\usepackage[rm]{subfigure}   
\usepackage{threeparttable}
\usepackage[utf8]{inputenc}
\usepackage[T1]{fontenc}
\usepackage{algorithm}
\usepackage{multirow}
\usepackage[normalem]{ulem}
\usepackage{float}
\usepackage{flafter}
\usepackage{algorithmic}

\usepackage{amsmath}

\usepackage{algorithm}
\usepackage{indentfirst}
\usepackage{setspace}

\usepackage{tikz}
\definecolor{gold}{rgb}{0.85,.66,0}

\begin{document}
%==============================================================
\title{Hybrid Hughes-Hartogs Power Allocation Algorithms for OFDMA Systems}
%==============================================================
\author{Joao Henrique Inacio de Souza, Taufik Abrao
\thanks{Electrical Engineering Department, State University of Londrina.  Rod. Celso Garcia Cid - PR445, Po.Box 10.011. CEP:86057-970, Londrina, PR  - Brazil. E-mail: \texttt{taufik@uel.br \quad joaohis@outlook.com}}
\thanks{``This work was supported in part by the  National Council for Scientific and Technological Development (CNPq) of Brazil under Grants 304066/2015-0; by the Londrina State University (UEL) and the Paran\'a State Government''}}

\maketitle

\begin{abstract}
This work analyzes the discrete solution of Hughes-Hartogs (HH) for the transmission rate maximization problem with power constraint in the OFDMA systems and explores mechanisms to reduce the computational complexity of greedy algorithms. In addition to the solution characterization, a computational complexity analysis is developed, considering the number of executed operations for running time purpose.  Moreover, we have compared the system capacity via the throughput obtained with the HH solution, and its variants combined with three complexity reduction mechanisms.  These tools consist of an initial allocation bit vector calculated by rounding the results of the water-filling (WF) solution, the multiple subchannels per iteration updating, and the adoption of a subchannel grouping procedure. Our findings indicate that the update of multiple subchannels and the subcarriers grouping techniques reduce the number of iterations required for convergence of the original HH, with some throughput degradation. Also, the bit-allocation mechanism based on the WF is deployed as an alternative to overcome the HH solution, increasing the computational complexity.
\end{abstract}


%%%%%%%%%%%%%%%%%%%%%%%%%%%%%%%%%%%%%%%%%%%%%%%%%%%%%%%%
\section{Introduction}\label{sec:intro}
%%%%%%%%%%%%%%%%%%%%%%%%%%%%%%%%%%%%%%%%%%%%%%%%%%%%%%%%
The orthogonal frequency division multiplexing  scheme is an alternative for wireless transmission among channels which suffering from deep fading. However, dividing the transmission band into $N$ narrow subchannels generates the optimization problem of choosing the suitable power on each subchannel that maximizes the transmission rate (optimization criterion).

The rate maximization problem with power constraints analyzed herein was extensively explored in past works, specially in the context of the wired discrete multitone (DMT) systems, as well as in the wireless orthogonal frequency division multiplexing access (OFDMA) systems. Particularly, the OFDMA is based on the same principle of divide the total system bandwidth in shorter flat fading subchannels but sharing such spectral sub-bands {resources} with multiple users. 

{Associated} to these OFDMA features,  there is an involved {and intricated} resource allocation problem defined by the joint user subcarrier allocation and the respective subcarrier {optimal} power allocation. From different perspectives, the complete optimization allocating power, subcarriers, and bit-loading in multiuser OFDMA scenarios results in an exponential complexity to achieve optimality \cite{Tedik.2016, Fathi.2014, Xiong.2011, Kim.2005}.  As well known, the optimal solution for the power allocation problem is the water-filling solution \cite{shannon}, and optimally explored in \cite{gallager}. But this optimal solution assumes a non-integer number of allocated bits, which is inefficient on practical scenarios. In the other extreme, several sub-optimal jointly iterative methods, such as Dinkelbach, Lagrange dual decomposition algorithms, integer relaxation subcarriers allocation, subcarriers grouping mechanisms and so forth have been deployed (see \cite{Souza.2016} and related references inside) aiming at obtaining implementable resource allocation procedures.

Searching for practical solutions, discrete algorithms classified as bit-loading algorithms have been developed in the last decades. The optimal bit-loading solution is the Hughes-Hartogs (HH) algorithm \cite{hhpatent}, originally proposed for DMT systems. Because it is formulated on an exhaustive search procedure, the Hughes-Hartogs solution demands a lot of computational resources to be implemented. To minimize the computational complexity on discrete bit allocation, works such as \cite{cioffi_efficient} have proposed efficient solutions exploring the system margin to compute the result with less iterations.

Besides the rate maximization with power constraint, other works have studied resource allocation problems with constraints of different nature. Past works such as \cite{multiuser_adap}, \cite{hh_based_alg} have proposed solutions for the power allocation problem admitting simultaneous constraints, including power, bit error rate (BER) and modulation order.

The authors in \cite{sampaio} present a resource allocation algorithm {based on a non-cooperative game} to distribute power and subcarriers in {an} OFDMA LS-MIMO system, aiming to maximize the spectral efficiency of {an} inter-cell interference environment. {The found solution combines a} greedy procedure for subcarriers allocation and the classical water-filling solution to find the Nash equilibrium within a few iterations.

The authors in \cite{farzamnia} use a hybrid solution of the water-filling algorithm and a Nash game formulation to distribute the power on the subchannels of a MIMO-OFDM scheme. The proposed algorithm, which generalizes the applicability of the classical water-filling, jointly evaluate the power required to attend a target SNR value and efficiently distribute it among the subcarriers.

In the scope of cognitive radio (CR) systems, the work \cite{mohammadi} uses the convex optimization framework to maximize the throughput of a MIMO-OFDMA based scheme, maintaining a reasonable interference level on the primary users and subjected to a maximum power constraint. The authors split the original optimization problem into two coupled subproblems with more efficient solutions. Moreover, the authors in \cite{cao}  propose an algorithm for throughput maximization of OFDMA-MIMO-CR systems that includes subcarrier assignment, spatial beamforming, power allocation and bit-loading.

There are works proposing mechanisms for the bit-loading complexity reduction based on greedy algorithms, at the cost of throughput system degradation. For instance,  the work \cite{papandreou} presents a solution for DMT applications which applies the standard bit-filling and bit-loading algorithms combined with an initial bit-vector.  The solution is calculated by rounding the results of the WF optimal solution.  Moreover, in the power line communications (PLC) context,  beyond the initial bit-vector allocation, the authors in \cite{plc} propose a solution deploying techniques for computational complexity reduction, including a greedy algorithm updating multiple subchannels per iteration, {rather than just} one, while assuming computationally efficient approximations to compute the cost function.

The works \cite{nader}, \cite{asad} utilize subchannels grouping techniques to reduce the computational cost of the bit-loading solutions. The primary motivation of this kind of procedure comes from the naturally high correlation between the adjacent subchannels in an OFDM/OFDMA system operating under usual channel flatness condition. Hence, since the subchannel bandwidth is twice or further smaller than the channel coherence bandwidth, the channel state information (CSI) estimation of such adjacent subchannels can be grouped, and their respective channel gains can approximated by a fixed value, with a little loss of accuracy in such estimations.

The contributions of this work is twofold. We characterize the bit-loading problem in OFDMA systems with power constraint at the transmitter side, namely Hughes-Hartogs (HH). Moreover, we develop an extensive analysis on the average capacity {\it versus} complexity tradeoff for different bit-loading strategies, especially analyzing the HH grouping strategy. The complexity order of the algorithms and expressions for the complexity (running time) depending on the number of OFDMA subchannels, $N$, are determined. In addition, the performance of the HH algorithm is compared to its versions adopting three different mechanisms for the computational complexity reduction: {\it i}) the initial allocation bit vector using the WF solution, {\it ii}) the updating of multiple subchannels per iteration, and {\it iii}) the subcarriers grouping technique with different gain threshold by group.

The remainder sections of the paper are organized as follows. In Section \ref{sec:model}, we present the OFDMA system model considered, while in Section \ref{sec:problem} we state the power allocation problem and its discrete optimal solution. Afterwards, in Section \ref{sec:redcomplex} we present the three mechanisms to reduce the complexity of greedy algorithms, as well as the computational complexity of the HH algorithm is discussed in subsection \ref{sec:compcomplex}. After, in Section \ref{sec:numres} the numerical simulation results for the bit-loading power allocation algorithms are explored. Finally, the main conclusions are offered in Section \ref{sec:conclusion}.

% =====================================
\section{OFDMA System Model}\label{sec:model}
% =====================================
We consider an OFDMA system with $K$ users and $N$ subcarriers modulated by the QAM symbol vector:
\begin{equation}
	\mathbf{s}=[X[0],...,X[N-1]]^T,
\end{equation}
whose inverse discrete Fourier transforms (IDFT) is equal to
\begin{equation}
	\mathbf{x}=[x[0],...,x[N-1]]^T
\end{equation}

% É assumido que a largura de banda de cada subcanal é suficientemente menor que a banda de coerência do canal, i.e. $B<(\Delta f)_c$, sendo possível considerar que cada subcanal enfrenta desvanecimento plano. Também, assumiremos que o tempo de símbolo é inferior ao tempo de coerência do canal, i.e. $T_s<(\Delta T)_c$, e que um prefixo cíclico de duração maior ou igual ao máximo espalhamento temporal do canal é adicionado no início de cada símbolo, mitigando totalmente o efeito da interferência intersímbolica (ISI - \textit{intersymbol interference}). Para o processo de alocação de recursos, também será assumido que existe perfeita informação do estado do canal (CSI - \textit{channel state information}) durante o período de símbolo.

We assume the bandwidth of each subcarrier sufficiently narrower than the channel coherence bandwidth, \textit{i.e.} $B<(\Delta f)_c$, being possible to consider that each subcarrier is flat in frequency. In addition, assuming the symbol period duration less than the channel coherence period, {\it i.e.,} $T_s<(\Delta T)_c$, and admitting a cyclical prefix with time length equal or less than the channel maximum delay spread be added at the begin of each symbol, resulting the inter-symbol interference (ISI) effect could be completely mitigated. To perform the resource allocation process, we assume that there are perfect channel state information (CSI) feedback during each symbol period.

Considering $\mathbf{H}$ the circulant matrix with the channel impulse response, which is $\mu$ samples long, and the noise vector of each sample, respectively:
\begin{equation}
h[0],...,h[\mu], \qquad \text{and} \qquad \mathbf{n}=[n[0]~...~n[N-1]]^T,
\end{equation}
\begin{equation}
\mathbf{H}=\left[\begin{matrix}
h[0]&h[1]&\cdots&h[\mu-1]&0&\cdots&0\\
0&h[0]&\cdots&h[\mu-2]&h[\mu-1]&\cdots&0\\
\vdots&\ddots&\ddots&\ddots&\ddots&\ddots&\vdots\\
0&\cdots&0&h[0]&\cdots&h[\mu-2]&h[\mu-1]\\
\end{matrix}\right]
\end{equation}
\noindent suppressing the cyclical prefix samples, which suffer from ISI, the signal $\mathbf{y}$ on the receptor can be described as:
\begin{equation}
	\label{eq:recsig}
	\mathbf{y}=\mathbf{Hx}+\mathbf{n}
\end{equation}
%\colr{definir $\mathbf{H}$ e associar a $h[\mu]$ e/ou $\bf h$.}\\

Considering an $N\times N$ DFT $\mathbf{Q}$ matrix, the demodulated symbols at the receptor can be described as:
\begin{align}
	\mathbf{r}&=\mathbf{Qy} \, =\, \mathbf{QHx}+\mathbf{Qn}\nonumber\\
	&=\mathbf{QH}\mathbf{Q}^H\mathbf{s}+\mathbf{Qn}\nonumber\\
	&=\mathbf{\hat{H}s}+\mathbf{N}%\nonumber
\end{align}
\noindent where $\mathbf{N}$ is the DFT matrix of the noise vector $\mathbf{n}$, and $\mathbf{\hat{H}}$ is the diagonal matrix filled with the samples of the channel frequency response.

\noindent{\it Remark {\bf 1}.}\,\, We use $\mathcal{A}_k$ to denote the set the subcarriers assigned to the user $k$. In our work, we consider the orthogonal assignment of the frequency resources, {\it i.e.} each subcarrier can be allocated for a single user per OFDMA symbol. However, other schemes with the OFDMA characteristics allow subcarriers sharing by more than one user, {\it e.g.}, the MC-CDMA-MIMO model system treated in \cite{arumugam}.

Considering that {the focus of this work is} on the power allocation problem,  {herein, we will not consider the optimization of the OFDMA channel} resources distribution. So, we {have adopted} the criterion of dividing equally blocks of contiguous subcarriers between the users for subchannels assignment, such as the 3GPP LTE standard \cite{3GPPEVOLUTIONWP}. Other methods can be used for the subcarriers distribution, such as the use of a solver to evaluate the original mixed-integer optimization problem or its linear formulation presented on \cite{Inhyoung}, iterative sub-optimal algorithms \cite{Souza.2016} or heuristic solutions \cite{Kim}.

%%%%%%%%%%%%%%%%%%%%%%%%%%%%%%%%%%%%%%%%%%%%%%%%%%%%%%%%
% \section{Hybrid WFHH Solution}\label{sec:hybrid}
\section{Optimization Problem}\label{sec:problem}
%%%%%%%%%%%%%%%%%%%%%%%%%%%%%%%%%%%%%%%%%%%%%%%%%%%%%%%%
The classical resource optimization problem considered consists of the maximization of the total capacity of the OFDMA system with power constraint. The optimization problem is defined as follows:
\begin{align}\label{eq:problem}
\mathop{\rm maximize}\limits_{{\bf p}\in \Re_+^N}\quad & \sum_{k=1}^K\sum_{i\in\mathcal{A}_k}b_i=\sum_{k=1}^K\sum_{i\in\mathcal{A}_k}\log_2\left(1+\frac{p_{k,i}\delta_{k,i}}{\Gamma}\right) \\
	\rm{s.t.}\quad & %\sum_{k=1}^K
	 \sum_{i\in\mathcal{A}_k}p_{k,i}	 \leq P_{\max},  \quad k=1,2,\ldots,K\notag\\
	 &{\bf p} =[p_1, \, p_2, \ldots, \, p_N] \succeq {\bf 0}\notag
\end{align} 
where $N$ the number of OFDMA subchannels, $\mathcal{A}_k$ denotes the set of subcarriers assigned to the user $k$;
\begin{equation}
	\delta_{k,i}=\frac{|h_{k,i}|^2	}{N_0B}
\end{equation}
is the channel gain normalized by the thermal noise power, with $|h_{k,i}|^2$ {as} the power gain of the $i$-th subchannel for the user $k$, $N_0$ is the noise spectral density and $B$ is the band of each subchannel; $b_i$ and $p_i$ represent the number of bits and the power allocated on each subchannel, respectively, while $\Gamma$ the SNR gap and $P_{\max}$ the maximal available power constraint per user.

The optimal power vector solution for this problem is the well-known water-filling, which results in fractional allocated-bits because of its continuous characteristic. When we treat a practical scenario, the water-filling solution can't be implemented completely, because we would need infinite granularity on the system modulation constellation \cite{cioffi_wf}.

%...................................................
\subsection{HH Algorithm}
%...................................................
To overcome the granularity problem one can use a sub-optimal solution which assumes discrete bit allocation. Hence, when considering practical scenarios, the Hughes-Hartogs algorithm \cite{hhpatent} provides optimal solution for the optimization problem \eqref{eq:problem}, assuming discrete bits allocation, but sub-optimal solution when considering continuous information quantities. Moreover, the HH approach is based on exhaustive search, demanding a lot of computational resources to be computed.

\noindent{\it Remark {\bf 2}.}\,\, For the sake of notation simplicity, hereafter we have dropped the index identifying the $k$-th user, assuming that subcarrier allocation procedure has been done in a previous step, since herein our focus is on subcarrier power and bit-loading allocation. Also, for uniformity purpose, we have assumed that for each active user ($k=1,\ldots K$) in the OFDMA system there is an equal spectral resource available, defined by $N\cdot B$ [Hz].

The HH algorithm calculates the energy incremental cost to allocate one more bit on the subcarriers, choosing the one with the smallest cost \cite{hhpatent}. The process is performed bit-by-bit until the allocated energy reaches the constraint. The energy incremental cost to allocate the bit of number $b_i$ on the i-th subcarrier is equal to:
\begin{equation}
	\Delta\epsilon_i(b_i) = 2^{b_i}\frac{\Gamma}{\delta_i}.
\end{equation}
where $\delta_i$ is the channel gain normalized by the thermal noise power related to the user allocated on the $i$-th subcarrier.

Algorithm \ref{alg:hh} shows a pseudo-code for the HH solution. Firstly, it is defined the incremental cost matrix with the amount of energy required to allocate the first bit on each subchannel; $P_{\rm alloc}$ is the power sum allocated to the user in the current iteration. Afterwards, is executed a minimum search, allocating the bits on the channels whose present the lowest cost, until the total energy allocated reaches the power constraint.

\begin{algorithm}[H]
	\caption{HH bit-loading algorithm}\label{alg:hh}
	\begin{algorithmic}[1]
		\STATE \textit{{Initiate}}

		\STATE $P_{\rm alloc}=0$
		\STATE $b_i=0, \forall i=1,...,N$
		\STATE $\Delta\epsilon_i=2^{b_i}\frac{\Gamma}{\delta_i} \forall i=1,...,N$
		
		\WHILE{$P_{\rm alloc}<P_{\max}$}
		\STATE Find $c$ such that $\Delta\epsilon_{c}\leq\Delta\epsilon_i~\forall i \neq c$
		\STATE If $P_{\rm alloc}+\Delta\epsilon_{c}>P_{\max}$, terminate the algorithm
		\STATE Allocate 1 bit on the subchannel $c$
		\STATE $P_{\rm alloc}=P_{\rm alloc}+\Delta\epsilon_{c}$
		\STATE $\Delta\epsilon_{c}=2^{b_{c}}\frac{\Gamma}{\delta_{c}}$
		\ENDWHILE

		\STATE \textit{Terminate}
	\end{algorithmic}
\end{algorithm}

% =====================================
\section{Mechanisms for Complexity Reduction}\label{sec:redcomplex}
% =====================================
%...................................................
\subsection{Initial allocation bit vector}
%...................................................
In its original formulation, the HH algorithm initiates with the bit vector null, or filled with the maximum number of bits allowed on the system, dependent on the adoption of the bit-filling or the bit-removal criteria.

The choose of {a} specific initial allocation bit profile can reduce the iterations number required for convergence of the allocation algorithm, decreasing its running time. Some works study the initial allocation bit vector that minimizes the iterations number of the system.

One possible initial allocation bit vector is calculated by rounding the results of the WF solution, evaluating discrete results. The WF solution is developed applying the Lagrange multipliers optimization technique on the problem \eqref{eq:problem}, resulting in the equation:
\begin{equation}
	\label{eq:wfpow}
	p_i=C_\lambda-\frac{\Gamma}{\delta_i}
\end{equation}
where $C_\lambda$ is a constant called "water level":
\begin{equation}
	\label{eq:wfwat}
	C_\lambda=\frac{1}{N}\left(P_{\max}+\sum_{i=1}^{N}\frac{\Gamma}{\delta_i}\right)
\end{equation}

With the vector $\mathbf{p^{(0)}}=[p^{(0)}_1~...~p^{(0)}_{N}]$ calculated by the equation \eqref{eq:wfpow}, the elements $b^{(0)}_i$ of the initial allocation bit vector $\mathbf{b^{(0)}}=[b^{(0)}_1~...~b^{(0)}_{N}]$ are evaluated by the expression:
\begin{equation}
	\label{eq:bitinit}
	b^{(0)}_i=\Bigg\lfloor\log_2\Bigg(1+\frac{p^{(0)}_i\delta_i}{\Gamma}\Bigg)\Bigg\rfloor
\end{equation}
Thus, the allocation process initiates with the initial allocation bit profile defined by the $\mathbf{b^{(0)}}$ obtained vector, {rather than} the prior null bit vector.

%...................................................
\subsection{Update multiple subcarriers per iteration}
%...................................................
The HH algorithm implementation given on the pseudo-code Algorithm \ref{alg:hh} predicts the update of only one bit on only one subcarrier per iteration. One alternative for operation consists on updating only one bit on $\kappa$ subcarriers during each iteration, choosing the $\kappa$ subcarriers with the lowest bit incremental energy cost.

The process of updating multiple subcarriers is a simple strategy to reduce the iterations number required for convergence; at prior, the mechanism is able to decrease the number of iterations of the HH algorithm by a $\kappa$ factor. Despite that, this strategy results in degraded capacity rates in comparison with the obtained by the HH algorithm in its original formulation, because these strategy don't guarantee the minimal incremental cost at each algorithm step. For example, at a determined step in an algorithm execution with $N=2$ and $\kappa=2$, allocate 2 bits on the subcarrier 1 may have smaller incremental cost than dividing the 2 bits between the subchannels.

%The process of updating multiple subcarriers is a simple strategy to reduce the iterations number required for convergence; at prior, the mechanism is able to decrease the number of iterations of the HH algorithm by a $\kappa$ factor. Despite that, this strategy results in degraded capacity rates in comparison with the obtained by the HH algorithm in its original formulation, once the update of only one bit on each $\kappa$ subcarrier with the more favorable conditions in the incremental cost point of view does not guarantee that this step consists on the system local optimum under its conditions.

%...................................................
\subsection{Subcarriers grouping strategy}\label{ssec:grp}
%...................................................
The subcarriers grouping process has the objective of divide the system's subchannels in sets with fixed channel gain. This process can be done using grouping algorithms, generally based on the subcarriers' gain. As a consequence of the grouping process, the subcarriers are reduced to their groups, performing the allocation process onto the groups, in place of each subcarrier independently.

The channel impulsive response $r(t)$ can be modeled as the sum of $L$ delayed multipath components,
\begin{equation}
	r(t)=\sum_{i=1}^La_i\delta (t-\tau_i)
\end{equation}
where $a_i\sim\mathcal{CN}(0,\sigma_{a_i}^2)$ is the $i$-th multipath amplitude at the receiver, $\tau_i$ is the $i$-th multipath delay dependent on the environment's scatter geometry and $\sum_{i=1}^L\sigma_{a_i}^2=1$. Evaluating the $N$-points discrete Fourier transform (DFT) of $r(t)$, we obtain the channel coefficient of the $n$-th OFDM subchannel as
\begin{equation}
	R_n=\sum_{i=1}^La_ie^{-j2\pi f_n\tau_i}, \qquad n=1,\ldots,N
\end{equation}
where $f_n$ is the $n$-th  {subchannel central} frequency.

Evaluating the correlation coefficient $\rho_{n,m}$ between the $n$-th and $m$-th OFDM subchannels considering orthogonal multipaths, \textit{i.e.} $E[a_ia_l^*]=0,~\forall i\neq l$,
\begin{align}
	\rho_{n,m}=& \mathbb{E}[R_nR_m^*] = \mathbb{E}\Bigg[\sum_{i=1}^L\sum_{l=1}^La_ia_l^*e^{j2\pi (f_m\tau_l-f_n\tau_i)}\Bigg] \notag\\
	\rho_{n,m}=&\sum_{i=1}^L\sigma_{a_i}^2e^{j2\pi\tau_i(f_m-f_n)}\label{eq:corr}
\end{align}

\noindent{\it Remark {\bf 3}.}\,\, From \eqref{eq:corr} one can conclude that the correlation between the subchannels depends on their frequency separation ($|f_m-f_n|$), as well as the channel's power delay profile, which defines the $\tau_i$ values. Hence, the subcarrier grouping mechanism can be applied aiming at obtaining implementable low-complexity power allocation procedures for OFDM and OFDMA systems.

Considering specific system conditions, the subcarrier grouping procedure in OFDMA allows the approximation of the subchannel gains to a fixed value, with a small error due to the high correlation associated with the adjacent subcarriers. These conditions are related to the frequency gap between the subcarriers and the channel coherence bandwidth, which depends on the delay spread associated with the system channel impulse response. Channels with small delay spread presents more correlated subcarriers, when the opposite can be verified on channels with high delay spread. Hence, the performance of the subcarrier grouping process depends on the delay spread profile of the OFDMA channel.

\begin{algorithm}[H]
	\caption{Subcarriers Grouping Algorithm}\label{alg:grp}
	\begin{algorithmic}[1]
		\STATE \textit{{Initiate}:} $G_{\textsc{t}}$
		
		\STATE $c=1$
		\STATE $d=1$

		\FOR{$i=2,...,N$}
		\IF{$g_i \in [g_{c}-G_{\textsc{t}};g_{c}+G_{\textsc{t}}]$}
		\STATE Allocate the subchannel $i$ into the group $d$
		\ELSE
		\STATE $d=d+1$
		\STATE Allocate the subchannel $i$ into the group $d$
		\STATE $c=i$
		\ENDIF
		\ENDFOR

		\FOR{$i=1,...,d$}
		\STATE Define the gain of the group $i$ as the minimum of its subcarriers
		\ENDFOR

		\STATE \textit{Terminate}
	\end{algorithmic}
\end{algorithm}

The adopted subcarriers grouping algorithm is stated on the pseudo-code in the Algorithm \ref{alg:grp}. The algorithm initiates defining the parameter $G_{\textsc{t}}$, called gain threshold, in dB. After the definition of $G_{\textsc{t}}$, the first subchannel is declared as the leader of the first set. Thereafter, the adjacent subcarriers are scanned testing if their gain values belongs to the interval $[g_{c}-G_{\textsc{t}};g_{c}+G_{\textsc{t}}]$, where $g_{c}$ is the gain of the leader subchannel of the current set. If the current channel has the gain value inside the interval, the subcarrier is allocated into the same set of the leader subcarrier with index $c$. Otherwise, the current channel is allocated into a new group, and is defined as the leader of this group; the sequential grouping process of the OFDMA subcarriers is repeated until all the OFDM subcarriers are distributed into the groups. Subsequently the sequential grouping of the OFDMA subcarriers, the channel gain of each set is defined as the least between its elements.

% =====================================
\subsection{Computational complexity analysis}\label{sec:compcomplex}
% =====================================
The complexity analysis has the objective of identifying the resources number consumed by the algorithm, as hardware, memory and mainly running time, tracing a relationship between them and the obtained results at the end of the task. In general, the running time of an algorithm depends on its input length; it is common to describe the running time as a function of the input length. This analysis can be done counting step-by-step the operations executed on each line code of the pseudo-code \cite{intro_alg}.

The basic arithmetic operations are implemented in hardware with fixed computational cost, not influenced by the input length. Although more complex algorithms like sort operations have a running time proportional to the input length. So we need to identify these types of operations and consider their particular running time.

To obtain a general complexity analysis of the algorithm we need to count the executed operations on its worst case execution. After counting, the obtained expression of the running time dependent on the input length is a superior bound for the running time.

%To analyze the running time of the HH algorithm, we needed to calculate the running time of each instruction executed by it. We did this identifying and counting on the code the number of basic arithmetic operations and more complex operations, like sort and search.

The running time of the HH algorithm can be calculated summing the execution time of each instruction performed by it. Therefore, we have to identify and count the number of basic arithmetic operations and more complex operations, such as sort and search, on the algorithm code. It is important to identify the operations which are executed repeatedly inside the loops, because their running time depends on the length of the input data, being the more significant instructions on the total algorithm running time. After the counting, the time contributions of each instruction are summed, obtaining the total running time expression.

First, we counted the basic arithmetic operations: addition, subtraction, multiplication, division, exponents and logarithms. During counting, we divided the operations into two groups: a) group 1 for the operations processed once on the algorithm; b) group 2 of operations processed repeatedly with the algorithm iterations.

The Tab. \ref{tab:num_operat} presents the number of basic arithmetic operations classified into the group 1 and the group 2. After the counting, we determined the complex operations. We identified one sort inside the algorithm's loop. According to \cite{intro_alg}, this operation can be implemented with algorithms that have the execution time upper bounded by the function $u(N)=N\log(N)$.

\begin{table}[!b]
\caption{Number of operations executed in the algorithm HH once (group 1) and repeatdly inside the loop (group 2).\label{tab:num_operat}}
\centering \small
\begin{tabular}{ccc}
\hline
\textbf{Operation} & \textbf{Group 1} & \textbf{Group 2}\\
\hline
			Addition		&	0	&	$2N+1$\\
			Subtraction		&	0	&	0\\
			Multiplication	&	$N$	&	1\\
			Division		&	0	&	0\\
			Exponent		&	$N$	&	1\\
			Logarithm		&	0	&	0\\
			\hline
\end{tabular}
\end{table}

An numerical analysis of the iterations number of the HH algorithm as a function of the system's subcarriers number $N$ indicates a linear dependence. Therefore, assuming that the iterations number is a function $v(N)$ which depends on the subcarriers number $N$, for the algorithm HH we have $v_{\textsc{hh}}(N)=N$.

Hence, summing the contributions of all the HH algorithm operations, we obtain the following expression for its running time:
\begin{equation}
	\label{eq:hybrid_running_time}
	\mathcal{T}_\textsc{hh}(N) = N^2\log{N}+2N^2+5N
\end{equation}
Expressing the calculated $\mathcal{T}_\textsc{hh}(N)$ in the notation of the asymptotic superior bound $\mathcal{O}\{.\}$, we have that the HH algorithm complexity is bounded by:
\begin{equation}
\mathcal{T}_\textsc{hh}^{\text{asym}}(N)=\mathcal{O}\{\mathcal{T}_\textsc{hh}(N)\}=N^2\log{N}.
\end{equation}

% =====================================
\section{Numerical Results}\label{sec:numres}
% =====================================
With the objective of compare the computational complexity reduction and the capacity degradation verified with the adoption of the bit-loading simplified mechanisms, we implemented Monte-Carlo simulations with the allocation algorithm HH, and its versions using the initial allocation bit vector calculated by the WF algorithm (HH-WF), updating multiple subcarriers per iteration (HH-K), and with a subcarriers grouping algorithm (HH-GRP). Further the discrete allocation algorithms, we executed in the simulations the WF optimal solution and the uniform power allocation criteria (EQ), in order to compare the efficient but sub-optimal solutions with their respectively superior and inferior bounds for the capacity reached by efficient bit-loading algorithms.

The proposed simulations scenario consists on the downlink (DL) into an urban macro-cell with radius $r=1$ km and 8 users whose the positions are uniformly distributed into the cell. We considered a situation without line of sight (NLOS), where the received amplitudes in the receptor follow a Rayleigh statistical distribution. The subcarriers number was set from 128 to 4096 doubling the initial value; the total subchannels number was divided equally between the users. The Tab. \ref{tab:param} summarizes the values for the general system parameters and the employed channel.

\begin{table}[!b]
\caption{Parameters of the proposed simulations scenario.\label{tab:param}}
\centering \small
\begin{tabular}{cc}
			\hline
			\textbf{Parameter} & \textbf{Value}\\
			\hline
			Cellular system & OFDMA\\ 
			Macro-cell & $r=1000$ m\\
			Users number & $8$\\
			Subcarriers number & $N=128$ to $4096$\\
			Total bandwidth & $B=2$ MHz\\
			Path-loss exponent & $\xi = 4$\\
			NLOS channel fading & Rayleigh\\
			Gain threshold in the HH-GRP & $G_{\textsc{t}}\in \{\frac{1}{4}, \, \frac{1}{2}, \, 1;\, 5\}$ dB\\
			Subcarriers number of the HH-K& $\kappa\in\{2, \, 4,\, 8,\, 16\}$\\
			Power constraint & $P_{\max} = 10$ W\\
			Maximum bit error rate & BER$ =10^{-12}$\\
			Maximum delay spread & $\tau_{\max}=2.5~\mu$s\\
			\hline
\end{tabular}
\end{table}

In the proposed scenario we did $10^4$ Monte-Carlo realizations performing the power allocation process with the algorithms HH, HH-WF, HH-K with the $\kappa$ values equal to 2, 4, 8 and 16, and the HH-GRP with the $G_{\textsc{t}}$ values equal to 0.25 dB, 0.5 dB, 1 dB and 5 dB, WF and EQ. During each iteration we sorted 10 taps of the channel impulse response and the users' positions into the cell, evaluating the channel coefficients.

\begin{figure}[!htbp]
	\centering
	\includegraphics[width=.8\textwidth]{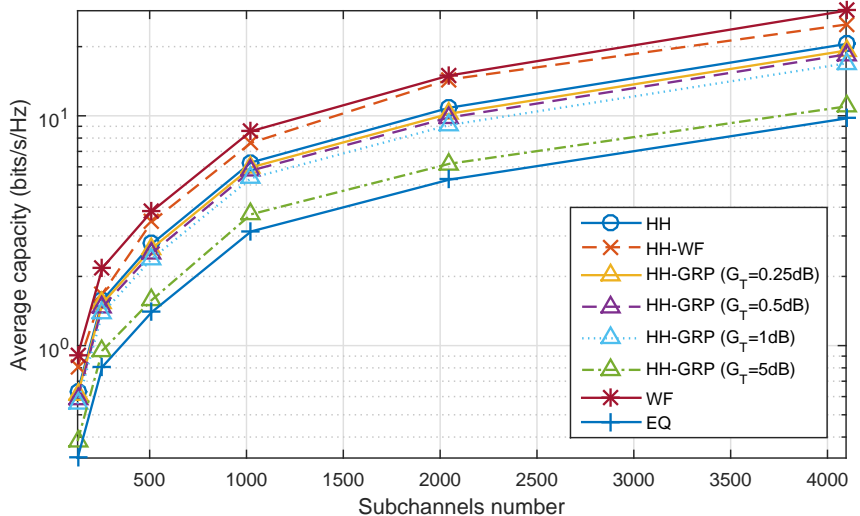}
	\vspace{-3mm}
	\caption{Average capacity as a function of the subcarriers number for the algorithms HH, HH-WF, HH-GRP, WF and EQ bit-loading algorithms with different gain thresholds.}
	\label{fig:capacity_hhgrp}
\end{figure}

The Fig. \ref{fig:capacity_hhgrp} presents the average capacity reached by the algorithms HH, HH-WF, HH-GRP, WF and EQ as a function of the subcarriers number. As the expected, the HH, HH-WF and HH-GRP algorithms reached average capacity values lower than the optimal solution WF, once the adoption of a discrete solution implies on lose of optimally. Among the discrete solutions, the HH-WF algorithm reached the highest capacity, overcoming the HH algorithm results. The HH-GRP algorithm results has shown average capacity values lower than the reached by the HH algorithm, with the degradation increasing with the grouping threshold $G_{\textsc{t}}$. This behavior is due to the reduction of the groups granularity increasing $G_{\textsc{t}}$, allowing the grouping algorithm to underestimate the subchannels, grouping them with other subchannels with lower gain.

\begin{figure}[!htbp]
	\centering
	\includegraphics[width=.82\textwidth]{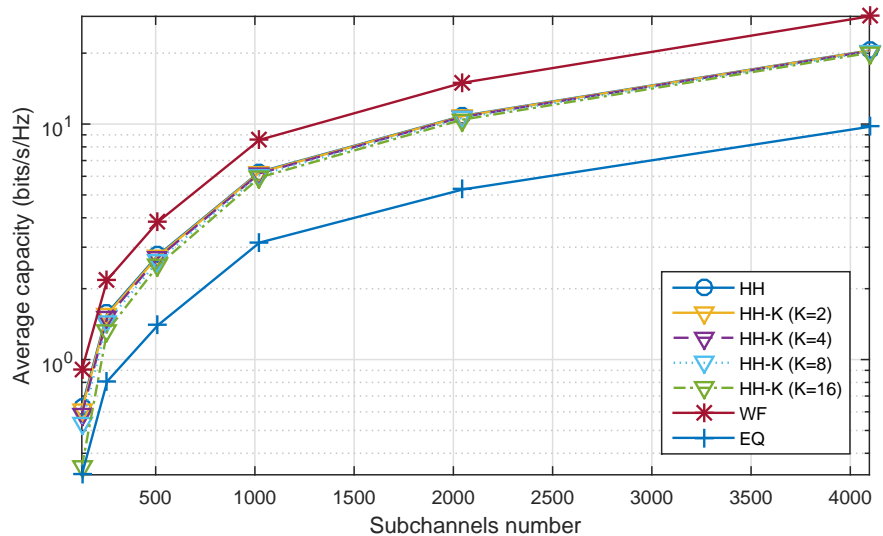}\\
	{\small a) HH, HH-K, WF and EQ}\\
	\includegraphics[width=.82\textwidth]{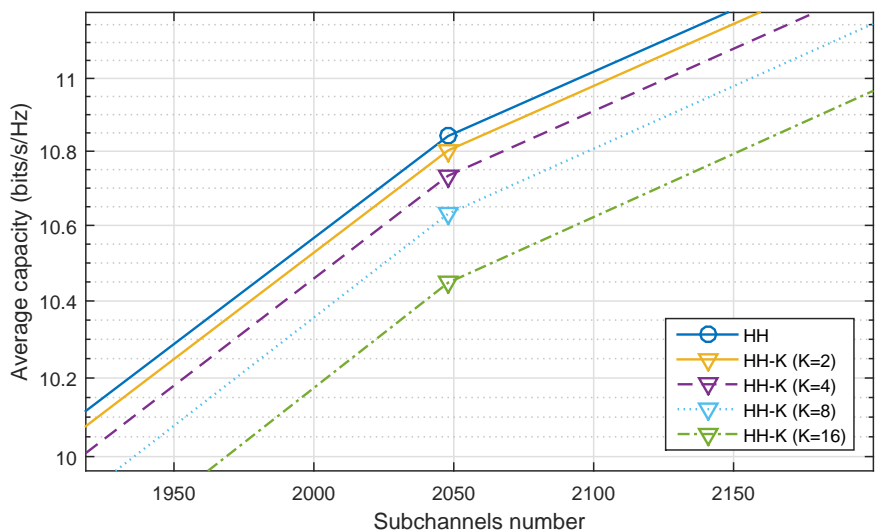}\\
	{\small b) HH and HH-K: zoom in $N=2048$ subcarriers}.
	\vspace{-2mm}
	\caption{Average capacity {\it versus} the number of subcarriers for the HH, HH-K, WF and EQ bit-loading algorithms}
	\label{fig:capacity_hhk1}
\end{figure}

Fig. \ref{fig:capacity_hhk1}.a) depicts the average capacity curves for the HH, HH-K, WF and EQ algorithms as a function of the subcarriers number. As the case of the HH-GRP algorithm, the capacity values reached by the HH-K one was lower than the calculated with the original HH. The growing of the $\kappa$ factor has reflected on the degradation increase in the capacity of the HH algorithm, as one can see on the zoom in the Fig. \ref{fig:capacity_hhk1}.b). The behavior is due to the fact of updating only one bit on the $\kappa$ more favorable subchannels during each iteration doesn't guarantee the minimal incremental cost at each algorithm step. Despite the degradation, one can see a marginal reduction of the capacity with the increase of the $\kappa$ factor when it is compared with the increase of the threshold $G_{\textsc{t}}$ in the HH-GRP algorithm shown in Fig. \ref{fig:capacity_hhgrp}.

\begin{figure}[!htbp]
	\centering
	\includegraphics[width=.82\textwidth]{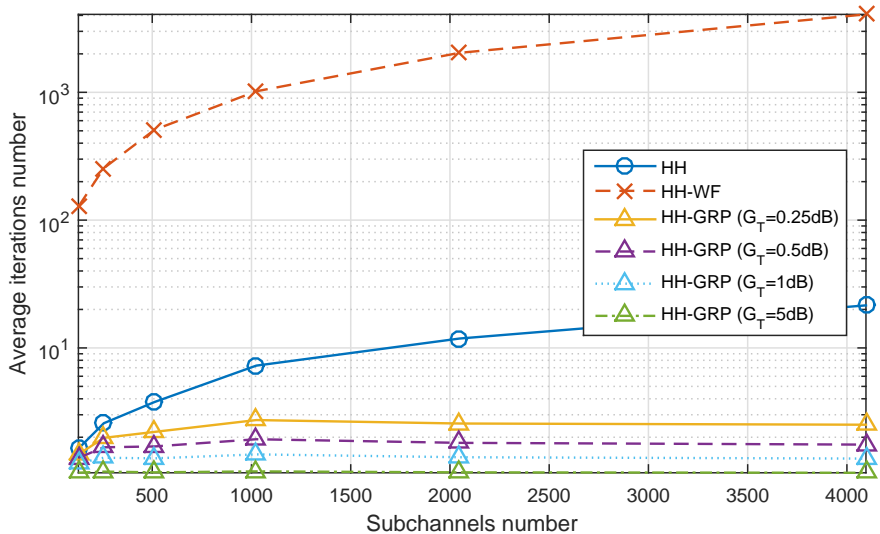}
	\vspace{-5mm}
	\caption{Average number of iterations required for convergence {\it versus} the subcarriers number for the algorithms HH, HH-WF and HH-GRP.}
	\label{fig:iterations_hhgrp}
\end{figure}

The Fig. \ref{fig:iterations_hhgrp} depicts the curves of the average iterations number of the algorithms HH, HH-WF and HH-GRP increasing the subcarriers number. The WF and EQ algorithms were suppressed on this analysis because they need a few iterations for convergence. As the expected, we noted that increasing the gain threshold $G_{\textsc{t}}$ decreases the iterations number required for convergence. This occurs because increasing $G_{\textsc{t}}$ reduces the number of groups, allowing the HH-GRP algorithm to operate with a search set whose the number of elements is less than the subcarriers value $N$. The algorithm HH-WF demands an iterations number higher (about two and four growing orders) than the other analyzed algorithms for convergence. Hence, the capacity gain, provided by the initial allocation bit vector calculate by the WF solution, compared to the original HH algorithm has a substantial complexity cost, according to the iterations number. 

\begin{figure}[!htbp]
	\centering
	\includegraphics[width=.82\textwidth]{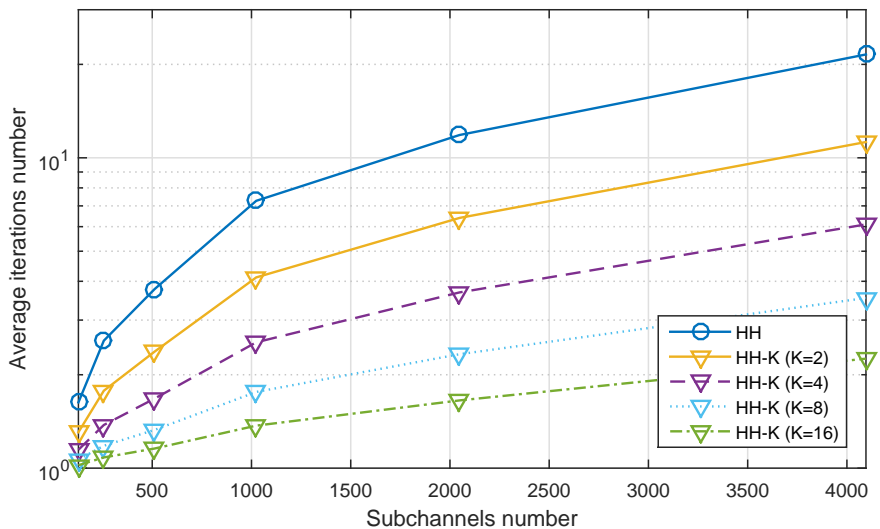}
	\vspace{-5mm}
	\caption{Average number of iterations required for convergence as a function of the subcarriers number for the algorithms HH and HH-K}
	\label{fig:iterations_hhk}
\end{figure}

To analyze the variation of the average iterations number of the algorithms HH-K and HH as a function of the subcarriers number, we trace the curves of the Fig. \ref{fig:iterations_hhk}. The approach of update multiple subcarriers per iteration (parameter $\kappa >1$) produced reduction in the iterations number required for convergence of the HH algorithm. We observed too that increasing the $\kappa$ number of updated subcarriers is proportional to the reduction in the average iterations number.

Comparing the solutions HH-K and HH-GRP, we note that the first offers as main advantage capacity values near to the reached by the HH algorithm, while the last one provides a faster convergence but with the capacity results more degraded. The initial allocation bit vector calculated by the HH-WF solution has shown the best capacity performance, overcoming the results of the HH algorithm and getting closer to the optimal results of the WF solution. However, the HH-WF algorithm is the one which required the highest iterations number for convergence, needing more computational resource to be executed.

\begin{figure}[!htbp]
	\centering
	\includegraphics[width=.82\textwidth]{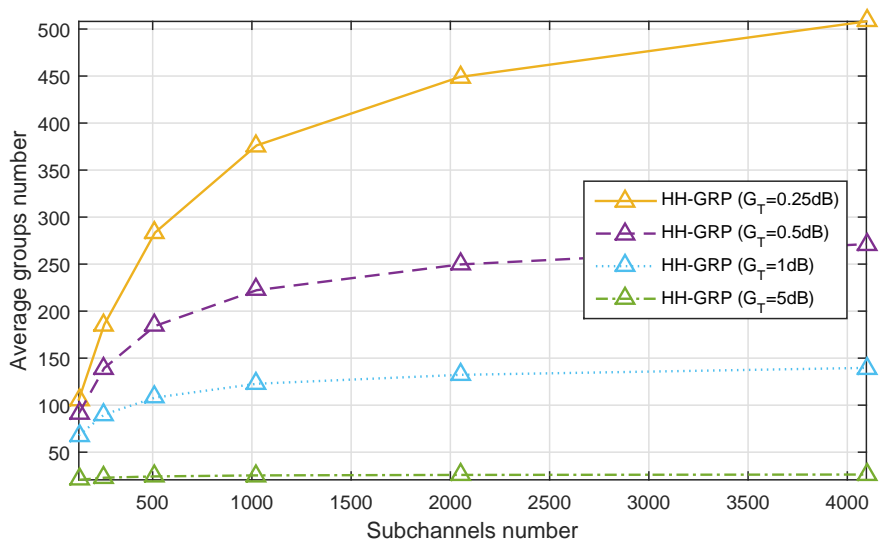}
	\caption{Groups number obtained by the grouping algorithm of the HH-GRP solution; $\tau_{\max}=2.5~\mu$s.}
	\label{fig:groups}
\end{figure}

To analyze the performance of the grouping algorithm HH-GRP with the change of the threshold gain $G_{\textsc{t}}$ and the subcarriers number of the system, we trace the curves of the average number of obtained groups by the subcarriers number, presented on the Fig. \ref{fig:groups}. The results confirm the reduction on the grouping granularity increasing $G_{\textsc{t}}$, implying on the reduction of the groups number. This is the major advantage of the technique, which reduce the elements number of the search set of the greedy algorithm, reducing the complexity for computation. Analyzing the behavior of the average groups number increasing the subcarriers number, we see an asymptotic behavior like roofs. This fact is due to the use of a channel model assuming a fixed maximum delay spread of the channel impulse response, {in this case, $\tau_{\max}=2.5~\mu$s}.

%-----------------------------
\subsection{Subcarriers Grouping Performance}
\begin{figure}[!htbp]
	\centering
	\includegraphics[width=.82\textwidth]{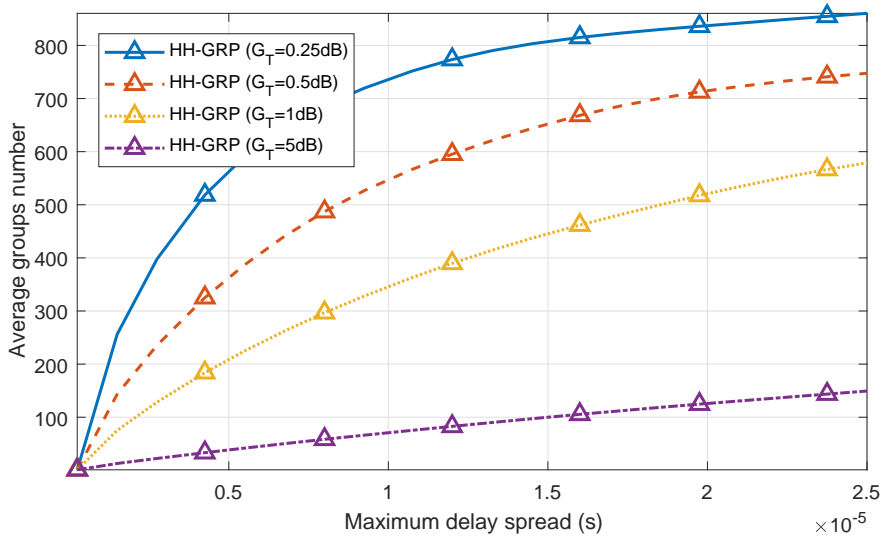}
	\vspace{-5mm}
	\caption{Average number of groups under different grouping thresholds for the HH-GRP algorithm solution as a function of the channel's maximum delay spread; $N=1024$ subchannels.}
	\label{fig:grp_delay_spread}
\end{figure}

%-----------------------------
{In this section, we analyze  the impact of the number of subcarrier groups on the HH-GRP performance under different channel conditions, including  channel delay spread $\tau_{\max}$ and OFDMA subchannels correlation}. Fig. \ref{fig:grp_delay_spread} depicts the curves of the average number of groups \textit{versus} the maximum delay spread, considering $N=1024$ subcarriers and different values for grouping threshold $G_{\textsc{t}}$. The increase of $\tau_{\max}$ conveys in more uncorrelated channels, increasing the number of groups, even for large SNR threshold values. This result supports the development of \eqref{eq:corr}, taking into account that the similarity between the channel's gain is the key role in the grouping algorithm.

\begin{figure}[!htbp]
\centering
	\includegraphics[width=.82\textwidth]{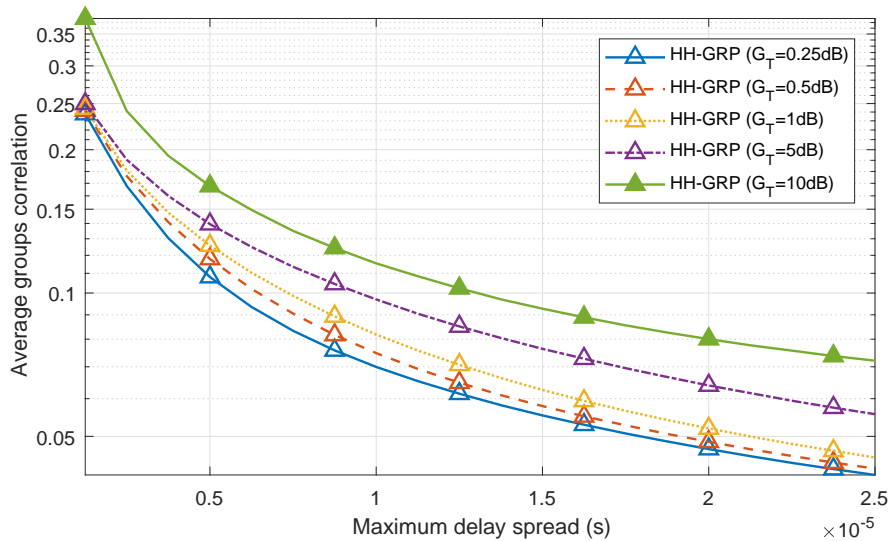}
	\vspace{-5mm}
	\caption{Average subcarrier-grouping correlation obtained by the grouping algorithm of the HH-GRP solution as a function of the channel's maximum delay spread; $N=1024$ subchannels.}
	\label{fig:grp_correlation}
\end{figure}

The normalized correlations between the different subcarrier groups formed by the HH-GRP grouping algorithm and varying the channel's maximum delay spread, are depicted in Fig. \ref{fig:grp_correlation}.  Such curves evidence that the average correlation coefficient of the subcarrier groups decreases very quickly when the delay spread increases {in the range $[1; \,\, 25]\mu$s}. We can see that, besides more uncorrelated subcarriers, the increase of $\tau_{\max}$ produces more uncorrelated groups. Moreover, the subcarrier groups correlation steadily decreases with the increment of SNR threshold granularity (small $G_{\textsc{t}}$).

\begin{table}[!htbp]
\caption{Summary of the numerical results for $N=1024$ subchannels.}\label{tab:num_res}
\small
\begin{tabular}{ccccccc}
\hline
\multicolumn{2}{c}{\textbf{Algorithm}} & \textbf{Avg. $\mathcal{C}^\star$} & \textbf{Avg. $\mathcal{I}^\star$} & \multicolumn{3}{c}{\textbf{Avg. number of groups}}\\
& & [bits/s/Hz] & & $\tau_{\max}=2.5~\mu$s & $\tau_{\max}=12.0~\mu$s & $\tau_{\max}=25.0~\mu$s\\
\hline
\multicolumn{2}{l}{WF} & 8.57 & - & - & - & -\\
\hline
\multicolumn{2}{l}{EQ} & 3.14 & - & - & - & -\\
\hline
\multicolumn{2}{l}{HH} & 6.26 & 7.23 & - & - & -\\
\hline
\multicolumn{2}{l}{HH-WF} & 7.63 & 1019 & - & - & -\\
\hline
\multirow{4}{*}{HH-K} & $\kappa=2$ & 6.23 & 4.12 & - & - & -\\
& $\kappa=4$ & 6.17 & 2.54 & - & - & -\\
& $\kappa=8$ & 6.08 & 1.76 & - & - & -\\
& $\kappa=16$ & 5.93 & 1.37 & - & - & -\\
\hline
\multirow{4}{*}{HH-GRP} & $G_T=0.25$ dB & 5.97 & 2.73 & 376 & 773 & 860\\
& $G_T=0.5$ dB & 5.78 & 1.93 & 222 & 595 & 748\\
& $G_T=1$ dB & 5.36 & 1.47 & 123 & 389 & 579\\
& $G_T=5$ dB & 3.73 & 1.08 & 25 & 83 & 149\\
\hline
\multicolumn{7}{l}{\small $^\star$Values calculated with $\tau_{\max}=2.5~\mu$s}
\end{tabular}
\end{table}

{Finally, the average capacity and the number of iterations required to perform the HH-GRP bit allocation have been determined by numerical  simulations assuming different conditions of channel's maximum delay spread and varying the gain threshold $G_T$, aiming to establish if there exists a $G_T$ value that maximizes the algorithm performance depending on the channel conditions. To compute the tradeoff between the capacity enhancement and the reduction of the number of iterations we have defined the tradeoff factor $\zeta\in [0;1]$ as follows
\begin{equation}\label{eq:tradeoff_factor}
\zeta=\frac{1}{2}\left(1+\frac{{\mathcal{C}}- \mathcal{C}_{\min}}{\mathcal{C}_{\max}-\mathcal{C}_{\min}}-\frac{{\mathcal{I}} -\mathcal{I}_{\min}}{\mathcal{I}_{\max}-\mathcal{I}_{\min}}\right)
\end{equation}
 where $\zeta=1$ indicate the best balance, $\mathcal{C}$ and $\mathcal{I}$ are, respectively, the average capacity and the average number of iterations, both calculated by numerical simulations, for each $\{\tau_{\max},G_T\}$-pair; moreover, the ratios  $\frac{\mathcal{C}_{\min}}{\mathcal{C}_{\max}}$ and $\frac{\mathcal{I}_{\min}}{\mathcal{I}_{\max}}$ are, respectively, the min-max capacity ratio and min-max number of iterations ratio calculated among all the scenarios evaluated. This equation {guarantees} that the {tradeoff} factor value represents equally the gain or loss of capacity and iterations number. Notice that in \eqref{eq:tradeoff_factor},  the capacity $\mathcal{C}$ and the number of iterations $\mathcal{I}$ are normalized, representing the same weight (50--50\%) on the  tradeoff factor calculation.} 

\begin{figure}[!htbp]
\centering
\includegraphics[width=.82\textwidth]{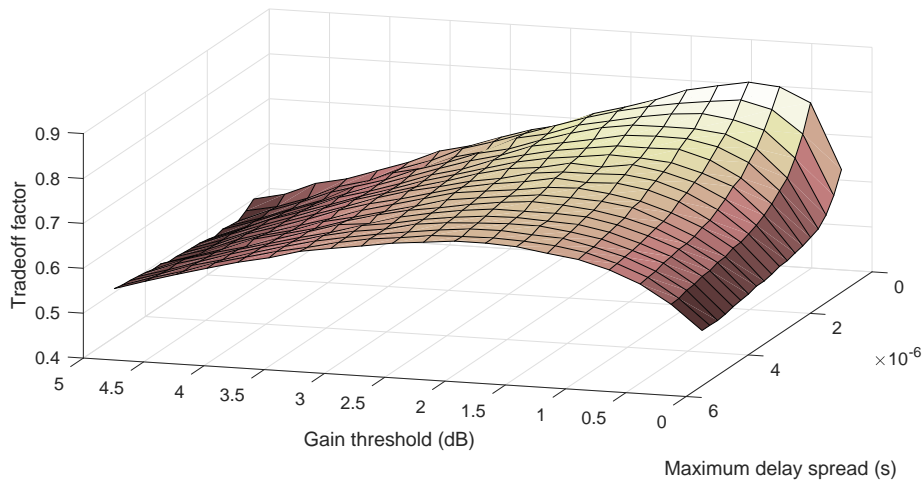}\\
{{\bf a}) $\zeta \times G_T \times \tau_{\max}$}\\
\includegraphics[width=.82\textwidth]{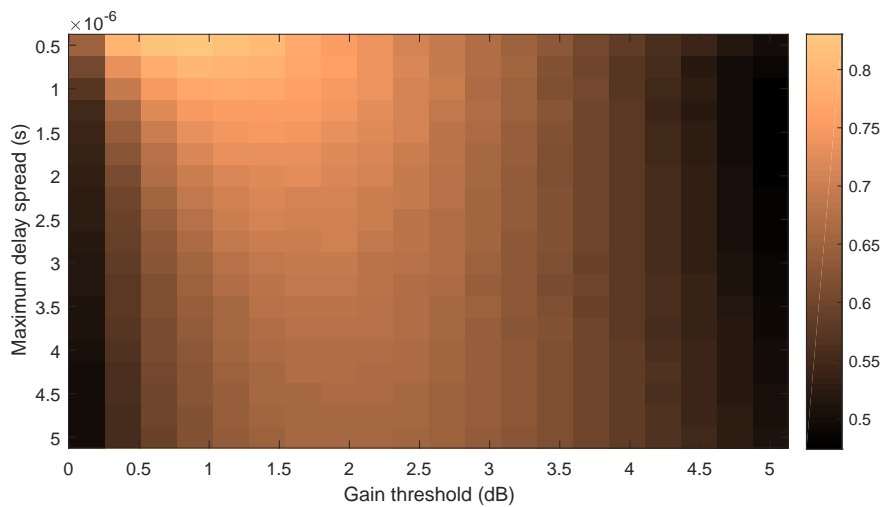}\\
{{\bf b}) Heat map for the same $\zeta \times G_T \times \tau_{\max}$}
\caption{{Tradeoff factor $\zeta$ calculated using the achieved capacity and the number of iterations for the HH-GRP algorithm under different conditions of the gain threshold $G_T$ and channel delay spread $\tau_{\max}$.}}
	\label{fig:compromise}
\end{figure}

{Fig. \ref{fig:compromise} depicts the tradeoff factor $\zeta$ for each pair of channel's maximum delay spread $\tau_{\max}$ and gain threshold $G_T$. The first consideration is that for all values of $\tau_{\max}$, choosing small or large values of $G_T$ results in low $\zeta$ factor values due to increasing the iterations number acts in favor of the capacity increasing (small $G_T$ values); alternatively, the capacity degradation works in favor of the decreasing complexity in terms of the number of iterations (large $G_T$ values), respectively.}

{Furthermore, for each value of channel's maximum delay spread, there exists a $G_T$ which offers the best tradeoff between the capacity enhancement and the reduction of the number of iterations required for convergence. This fact can be explained recalling the discussion of Section \ref{ssec:grp}, which shows that the correlation between adjacent subcarriers depends on the channel's delay spreading properties. Therefore, from the perspective of the wireless channel characteristics, the parameter $G_T$ can be adjusted in such a way that the HH-GRP algorithm is able to produce the best tradeoff between complexity and average capacity degradation.}

{Table \ref{tab:num_res} summarizes the numerical results of this section for capacity, average number of iterations and average number of groups for the HH-GRP algorithm, considering $N=1024$ subchannels and different values for $\tau_{\max}$ and the parameters of the HH-K and HH-GRP algorithms. As it was discussed earlier, the HH-K algorithm with $\kappa=16$ is the most efficient bit-loading solution, offering a remarkable reduction on the number of iterations when compared to the original HH algorithm, while resulting in a marginal degradation on the average capacity. Moreover, observing the average number of groups column, one can notice its dependence on the channel delay spreading and the grouping threshold parameters, whose reflects directly on the HH-GRP algorithm's performance.}

% =====================================
\section{Conclusion}\label{sec:conclusion}
% =====================================
The greedy Hughes-Hartogs algorithm for power allocation problem in OFDM systems was systematically characterized and compared with low-complexity sub-optimal approaches. Three mechanisms to reduce the algorithm computational complexity have been considered: a) the adoption of initial bit-vector allocation different from null bits profile; b) update multiple subcarriers per iteration; c)  subcarriers grouping technique.

The numerical simulations results evidenced the performance-complexity tradeoff of the analyzed bit-loading algorithms. The HH-GRP and HH-K solutions reached an average capacity lower than that attained by the original HH, needing a smaller iterations number for convergence. Comparing the HH-GRP and HH-K algorithms, one could observe that the HH-GRP results a higher reduction on the iterations number with more capacity degradation, being the two characteristics proportional to the gain threshold $G_{\textsc{t}}$, while the last one results in a much smaller capacity degradation, with the iterations number reduced proportionally to the factor $\kappa$. The HH-WF solution has shown the best average capacity results, overcoming the reached by the HH algorithm and getting closer to the calculated by the optimal solution WF. However, the HH-WF solution demands the highest iterations number for convergence, presenting the greatest computational cost. 

So, one could conclude that both mechanisms of updating multiple subchannels per iteration and the subcarriers grouping allow the convergence of the greedy algorithm HH with less iterations, at cost of some degradation on the average capacity. The algorithm with the best tradeoff on the performance {\it versus} complexity is the HH-K, which has demonstrated a low capacity degradation even with a high $\kappa$ value.

Finally, the initial bit-vector allocation defined by the WF solution was adopted as an alternative to obtaining an average capacity higher than the reached by the HH algorithm and its variants. As a consequence, a closer solution to the true WF one has been obtained, although this capacity gain comes with a computational complexity increasing, represented by the iterations number required for convergence.

% =====================================
\section*{Acknowledgment}
% =====================================
 This work was supported in part by the National Council for Scientific and Technological Development (CNPq) of Brazil under Grant  304066/2015-0, Fundacao Araucaria under Grant 302/2012, and in part by State University of Londrina -- Parana State Government (UEL) (scholarship).
 
%%%%%%%%%%%%%%%%%%%%%%%%%%%%%%%%%%%%%%%%%%%%%%%%%%%%%%%%
%%\bibliographystyle{iet}
%\bibliographystyle{IEEEtran}
%\bibliography{refHH}
%%%%%%%%%%%%%%%%%%%%%%%%%%%%%%%%%%%%%%%%%%%%%%%%%%%%%%%%

\end{document}